\documentclass[conference]{IEEEtran}

\ifCLASSINFOpdf
\else
\fi

\usepackage{amsmath}

\usepackage{url}

\usepackage{booktabs}
\usepackage{color}
\usepackage[hidelinks]{hyperref}
\usepackage{pifont}
\usepackage{tablefootnote}
\usepackage{tikz}
\usepackage{enumitem}
\usepackage{cite}

\usepackage{algorithm} 
\usepackage{algpseudocode}

\newcommand{\cmark}{\ding{51}}
\newcommand{\xmark}{\ding{55}}
\newcommand{\footnoteref}[1]{\hyperref[#1]{\textsuperscript{\ref*{#1}}}}

\newcommand*\circled[1]{\raisebox{.4pt}
                    {\tikz[baseline=(char.base)]{
            \node[shape=circle,draw,inner sep=1pt, style={fill=black, text=white}, scale=0.75] (char) {\textbf{#1}};}}}
 
\newcommand*\halfcirc[1][0.667ex]{%
  \begin{tikzpicture}
  \draw[fill] (0,0)-- (90:#1) arc (90:270:#1) -- cycle ;
  \draw (0,0) circle (#1);
  \end{tikzpicture}}

\newcommand{\para}[1]{\vspace{4pt}\subsubsection{\textbf{#1}}}

\begin{document}
\title{LightningSim: Fast and Accurate Trace-Based Simulation for High-Level Synthesis}

\author{\IEEEauthorblockN{Rishov Sarkar, Cong Hao}
\IEEEauthorblockA{School of Electrical and Computer Engineering, Georgia Institute of Technology\\
\href{mailto:rishov.sarkar@gatech.edu}{\nolinkurl{rishov.sarkar@gatech.edu}}, \href{mailto:callie.hao@ece.gatech.edu}{\nolinkurl{callie.hao@ece.gatech.edu}}}}

\maketitle

\begin{abstract}
High-Level Synthesis allows hardware designers to create complex RTL designs using C/C++. The traditional HLS workflow involves iterations of C/C++ simulation for partial functional verification and HLS synthesis for coarse timing estimates. However, neither C/C++ simulation nor HLS synthesis estimates can account for complex behaviors like FIFO interactions and pipeline stalls, thereby obscuring problems like deadlocks and latency overheads. Such problems are revealed only through C/RTL co-simulation, which is typically orders of magnitude slower than either C/C++ simulation or HLS synthesis, far too slow to integrate into the edit-run development cycle. Addressing this, we propose LightningSim, a fast simulation tool for HLS that combines the speed of native C/C++ with the accuracy of C/RTL co-simulation. LightningSim directly operates on the LLVM intermediate representation (IR) code and accurately simulates a hardware design's dynamic behavior.
First, it traces LLVM IR execution to capture the run-time information; second, it maps the static HLS scheduling information to the trace to simulate the dynamic behavior; third, it calculates stalls and deadlocks from inter-function interactions to get precise cycle counts. Evaluated on 33 benchmarks, LightningSim produces 99.9\%-accurate timing estimates up to 95\texttimes{} faster than RTL simulation. Our code is publicly available on GitHub.\footnote{\label{footnote:github}\url{https://github.com/sharc-lab/LightningSim}}
\end{abstract}

\IEEEpeerreviewmaketitle

\section{Introduction}

High-Level Synthesis (HLS) tools transform code in a high-level software programming language, typically C or C++ annotated with hardware-specific directives (e.g., ``pragmas''), into RTL code in Verilog or VHDL. 
Current HLS tools are capable of fast functionality verification through C/C++ simulation by compiling and executing the HLS source code. However, the latency estimation provided by HLS tools is usually far from accurate due to the lack of run-time information such as loop counts, branches, and stalls~\cite{flash,fastsim,wu2021ironman,sohrabizadeh2021enabling}. To get cycle-accurate timing (e.g., clock counts of operations, functions, and the entire program) and to check for design issues such as deadlocks, full-blown RTL simulation is needed, which is typically hundreds to thousands of times slower than C/C++ simulation and can take hours to days for complex designs.
How to get accurate latency information as fast as possible without running RTL simulation is essential to agile hardware development and is of great interest.

To address this challenge, we notice that, in order to generate RTL from C/C++ code, HLS tools must first compile the source code to LLVM~\cite{llvm} intermediate representation (IR) operations and precisely schedule them within functions and loops, which can already provide abundant (yet partial) latency information.
Latency variability can be attributed to the dynamic nature of the design, including factors such as loop iteration counts, branch conditions, and stalls in FIFOs and AXI interfaces.
Fortunately, \textit{such dynamic behaviors can be simulated through instrumentation of the C/C++ code only without invoking RTL simulation}. We can precisely infer the start cycle and cycle count of every LLVM IR instruction generated from the C/C++ code based on the \textit{IR execution trace}, to compute the accurate latency of the entire program.

\begin{figure}
    \begin{center}
        \includegraphics[width=\linewidth]{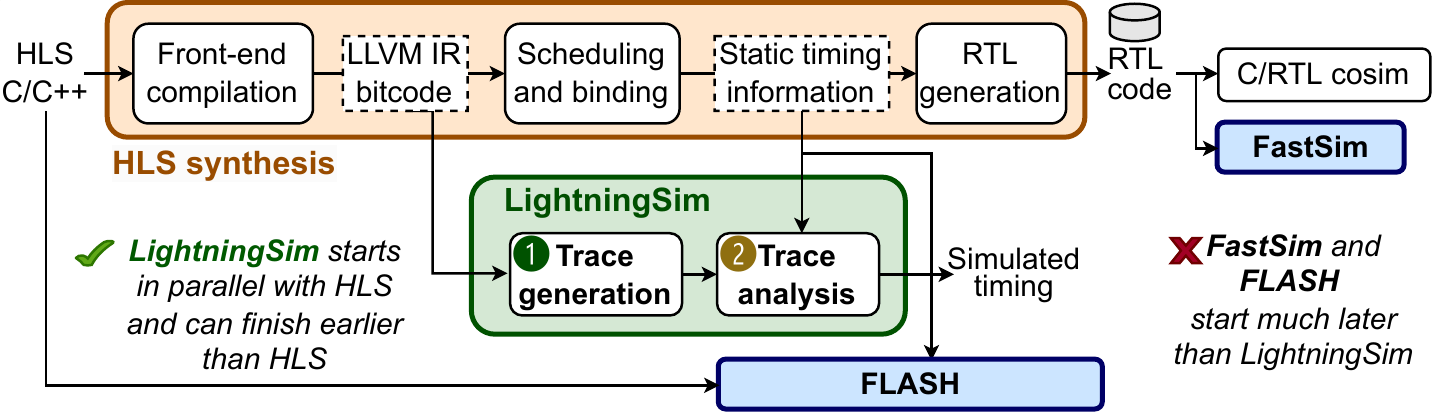}
    \end{center}
    \caption{LightningSim compared with HLS and existing simulation tools FLASH~\cite{flash} and FastSim~\cite{fastsim}. LightningSim can start in parallel with HLS and can finish earlier than HLS, while FastSim and FLASH start much later.}
    \label{fig:overall-with-HLS}
\end{figure}

Motivated by the need for faster simulation and the opportunities, we propose \textbf{LightningSim}, the first trace-based simulator based on Vitis HLS.\footnote{While LightningSim is based on Vitis HLS~\cite{vitis-hls} targeting FPGA, it should be easily applicable to other HLS tools with decent engineering effort.} It can provide accurate clock cycle counts for the entire program similar (99.9\%) to RTL simulation as soon as HLS finishes its front-end compilation and scheduling, even before RTL code generation.
Fig.~\ref{fig:overall-with-HLS} depicts its interactions with HLS and compares it with two existing fast simulation tools, FastSim~\cite{fastsim} and FLASH~\cite{flash}.
Notably, LightningSim can start in parallel with HLS as soon as the LLVM IR is generated after front-end compilation, while FLASH~\cite{flash} must wait till the scheduling is finished, and FastSim~\cite{fastsim} must wait till RTL generation finishes.

\begin{figure}
    \begin{center}
        \includegraphics[width=0.9\linewidth]{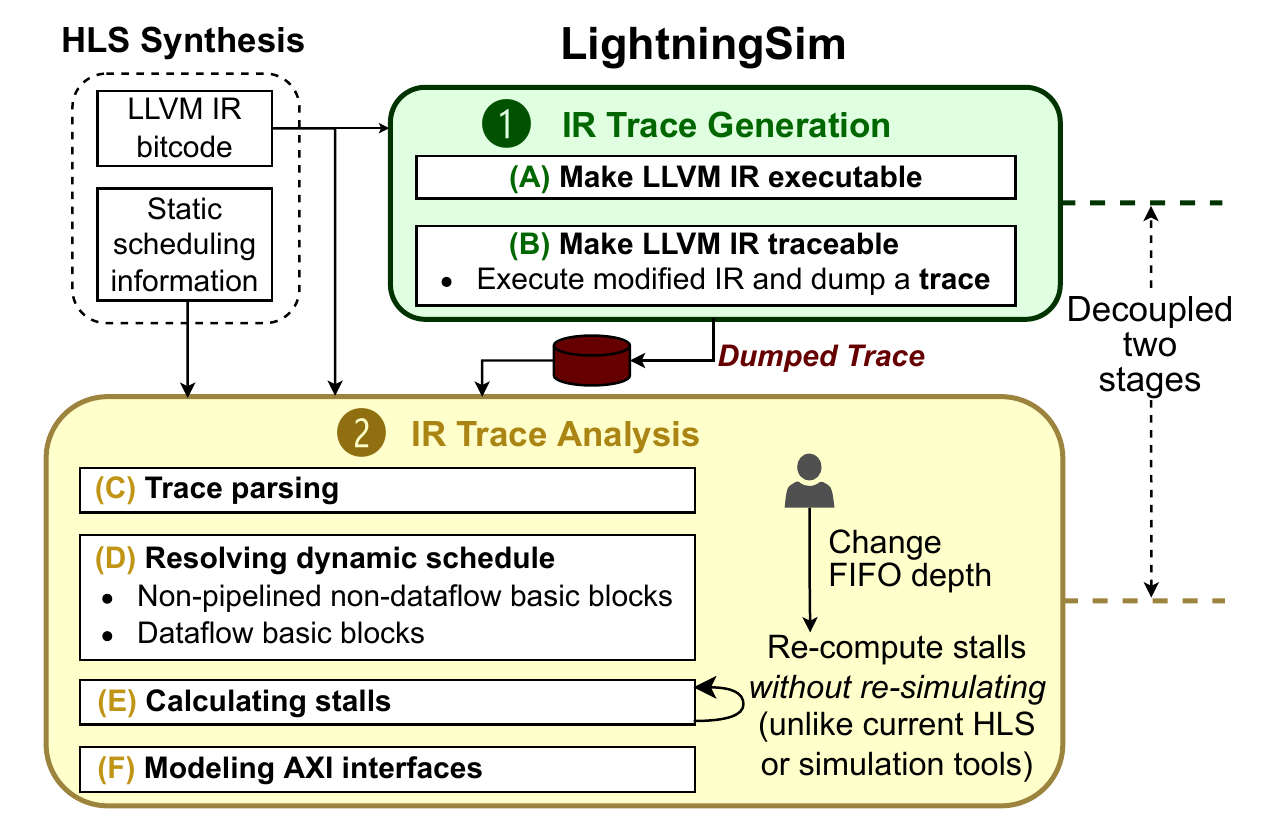}
    \end{center}
    \caption{An overview of LightningSim's decoupled two-stage simulation flow: IR trace generation and trace analysis.}
    \label{fig:LS-overview}
\end{figure}

In addition to being able to start early, the most significant advantage of LightningSim is a novel \textbf{decoupled} two-stage simulation: \circled{1} \textbf{IR trace generation} and \circled{2} \textbf{IR trace analysis}, as shown in Fig.~\ref{fig:LS-overview}.
\underline{First}, the IR \textit{trace generation} step takes in the generated LLVM IR by HLS, enables its execution and tracing by completing undefined functions on-the-fly, and then executes it on CPU to collect a \textit{trace}, i.e., a full execution history of LLVM IR instructions. \underline{Second}, the \textit{trace analysis} step maps the static scheduling information extracted from HLS onto the generated dynamic trace to calculate the resulting latency.
This step captures run-time and complex behaviors of the program, including pipelined loops, dataflow, FIFO/AXI stalls, branches, etc.
Decoupling the trace generation and analysis can enable \textit{incremental simulation}: if the hardware configuration, such as FIFO depth, is changed, it is only necessary to rerun trace analysis, rather than rerunning the entire HLS synthesis and trace generation. This is a unique advantage compared to existing HLS tools and previous simulators: if the simulation is based on the generated RTL, once the FIFO depth changes, then the entire RTL code needs to be re-generated and re-simulated.

As summarized in Table~\ref{tab:compare-existing}, we highlight our contributions and advantages over existing approaches as follows:
\begin{itemize}%
    \item \textbf{``C-like'' simulation speed with ``RTL-like'' accuracy.} LightningSim applies lightweight instrumentation to sequential LLVM code, which retains nearly the same performance as simply compiling and running the C/C++ program on a CPU. Moreover, LightningSim achieves almost the same cycle count as RTL simulation obtains.
    With 33 benchmarks, including state-of-the-art machine learning accelerators, LightningSim obtains 99.9\% accuracy in clock cycles compared to RTL simulation; the speedup is up to 95$\times$ with an average of 20.8$\times$.

    \item \textbf{Rigorous dynamic behavior modeling.} LightningSim faithfully mimics dynamic program behaviors, including FIFO stalls, deadlocks, pipelined loops, dataflow, branches, etc., using the static scheduling information provided by HLS and the dynamic IR execution trace.

    \item \textbf{Decoupled trace-based analysis for incremental simulation.} Thanks to decoupled trace generation and analysis, LightningSim can quickly and flexibly change simulated hardware parameters \textit{ex post facto}, after a single simulation run. For instance, to our best knowledge, LightningSim is the \textbf{first} tool that can simulate any FIFO depth from just one simulation, enabling LightningSim to modify or suggest optimal FIFO depths and detect deadlock without rerunning HLS synthesis.

    \item \textbf{Parallelizable with HLS synthesis.} To our best knowledge, LightningSim is the first tool that can start simulation before the completion of HLS synthesis or even scheduling. The \textit{IR trace generation} step can start as soon as the front-end LLVM compilation finishes, which happens before HLS scheduling, binding, and RTL generation. %
    In 13 out of 33 benchmarks, \textit{LightningSim finishes earlier than HLS synthesis}, especially for complicated benchmarks that require a long time for RTL generation.

    \item \textbf{Timing model for external memory accesses.} FPGA designs typically use AXI protocol to read and write from external memory, the DRAM. LightningSim includes a highly accurate timing model for such accesses.
    
    \item \textbf{Open-source with push-button ease of use.} LightningSim is publicly available on GitHub\footnoteref{footnote:github} and, following installation, one can easily apply it to existing Vitis HLS projects using a single Python command without rerunning HLS synthesis.
\end{itemize}

\section{Background and Motivation}

\newcommand{\limited}[1]{\hphantom{\textsuperscript{#1}}\halfcirc\textsuperscript{#1}}

\begin{table}
    \caption{LightningSim vs.\ existing  HLS software simulators.}
    \label{tab:compare-existing}
    \setlength{\tabcolsep}{3pt}
    \footnotesize
    \begin{center}
        \begin{tabular}{lccc}
            \toprule
            \textbf{Feature} & \textbf{FastSim~\cite{fastsim}} & \textbf{FLASH~\cite{flash}} & \textbf{LightningSim} \\
            \midrule
            Cycle-accurate for FIFOs & \cmark & \cmark & \cmark \\
            External memory modeling & \cmark & \xmark & \cmark \\
            
            Rapid incremental simulation & \xmark & \xmark & \cmark \\
            
            Parallelizable with HLS & \xmark & \limited1 & \cmark \\
            Event-driven simulation & \xmark & \xmark & \cmark \\
            
            Open-source & \limited2 & \xmark & \cmark \\
            \bottomrule
        \end{tabular}
    \end{center}
    
    {\footnotesize\textsuperscript{1}FLASH must wait till HLS scheduling finishes while LightningSim does not.}
    
    {\footnotesize\textsuperscript{2}At the time of writing, FastSim's public repository (\url{https://github.com/ckarfa/FastSim}) only contains pre-generated examples and no code for FastSim tool.}

\end{table}

\subsection{High-Level Synthesis Execution Flow and Limitations}

\noindent
\textbf{High-Level Synthesis execution flow.}
We use Xilinx Vitis HLS (formerly Vivado HLS)~\cite{vitis-hls} as a prototypical example, but the same concepts apply similarly to most other HLS tools.
HLS synthesis typically includes the following steps:

\paragraph{Front-end Compilation}
\label{sec:hls-compile}
HLS starts by compiling the input code into a lower-level intermediate representation, typically LLVM IR~\cite{llvm}. Optimization passes are performed on the IR, including custom HLS-specific passes to apply certain directives such as loop unrolling and array partitioning. %

\paragraph{Scheduling and Binding}
\label{sec:hls-schedule}
Each function in the LLVM IR is then scheduled and bound. During scheduling, HLS designs finite state machines (FSMs) along with pipelines and dataflows to control the execution of each LLVM instruction. During binding, HLS assigns hardware resources to implement the FSM and all instructions. %

\paragraph{RTL Generation}
\label{sec:hls-rtl}
Finally, HLS emits RTL code (Verilog and VHDL) and  reports for resource utilization estimates and rough timing estimates based on static analysis.

\noindent
\textbf{Limitations of High-Level Synthesis.}
Although HLS enables functionality verification by running fast C/C++ simulation, it does not provide accurate latency information with only static analysis.
As pointed out by previous studies~\cite{fastsim,flash,sohrabizadeh2021enabling}, HLS-generated latency reports can be far from accurate due to the lack of run-time information such as loop counts, branches, and FIFO stalls.
To obtain accurate latency, one can run ``C/RTL co-simulation'' (a.k.a. ``co-simulation'' or ``cosim''), in which the generated RTL code is simulated for cycle-accurate timing and deadlock detection. However, cosim is orders of magnitude slower than HLS synthesis, often taking hours to days, %
hindering the agility of the hardware design workflow.

\subsection{Existing Works and Key Insight}

We identify two prior works focusing on HLS simulation acceleration: FastSim~\cite{fastsim} and FLASH~\cite{flash}. %
FastSim translates the generated Verilog code back to equivalent C++ and optimizes the constructs that frequently appear such as finite state machines (FSMs) for faster simulation. FLASH uses the scheduling information extracted from HLS synthesis together with the input C/C++ code to generate a \textit{FIFO communication cycle-accurate} (FCCA) C model.
Both approaches rely on the information \textit{after} HLS scheduling or RTL generation.

By contrast, we identify an opportunity for a novel approach: \textbf{map HLS scheduling information directly onto LLVM IR execution.} The FLASH authors also commented on the possibility to simulate based on LLVM IR, though they chose C-level simulation for better readability~\cite{flash}. 
In this work, we advocate for IR-level simulation for the following reasons. \underline{First}, taking advantage of HLS front-end compilation, the LLVM IR already incorporates hardware information such as loop unroll and array partition, allowing for a closer simulation of actual hardware behavior, avoiding the need to repeat the effort made by HLS tools.
\underline{Second}, we can expose any differences in behavior between the original C/C++ code and the LLVM code, enabling more complete functional verification than C simulation alone. For example, our simulation accounts for nested loop flattening optimizations that cause variable definitions outside the inner loop to be hoisted inside of it.
\underline{Third}, since HLS scheduling information will be based on LLVM IR, we do not need to wait until scheduling finishes but can execute the IR as soon as it is generated. This also enables \textit{decoupled} execution, to be discussed shortly.
\underline{Fourth}, processing the IR makes LightningSim agnostic to high-level language syntax and can thus be more flexibly adapted to other HLS tools.

\subsection{On-the-Fly vs.\ Trace-Based Simulation}

While we propose mapping HLS scheduling information to LLVM IR, there are two viable approaches: \textit{one-step on-the-fly} simulation and \textit{decoupled trace-based} simulation.
LightningSim uses the latter, which plainly outperforms the former in two aspects: \textit{simulation start time}, and \textit{partial simulation when hardware configuration changes}.

\underline{First}, since \textit{on-the-fly} simulation executes the LLVM IR instructions and computes latency simultaneously using hardware scheduling information, the simulation cannot start until scheduling finishes. By contrast, in the \textit{decoupled trace-based} approach, trace generation can start as soon as HLS front-end compilation completes, which usually only takes a few seconds. 
\underline{Second}, since \textit{on-the-fly} simulation computes latency during execution, any tweaks to simulation parameters, such as FIFO depth, require restarting the entire simulation flow, including both recompiling the LLVM IR with new parameters and rerunning the compiled IR binary.
By contrast, the \textit{decoupled trace-based} approach does not need to rerun the generated trace in the first step, but can directly execute the second step to analyze the trace using new parameters. This is especially useful for designs with FIFOs and dataflow: LightningSim can decide optimal FIFO depth without repeatedly running HLS synthesis and cosim.

\section{LightningSim Overview and Challenges}

Fig.~\ref{fig:LS-overview} illustrates the overview of LightningSim, including two decoupled stages:
\circled{1} \textit{IR trace generation} and \circled{2} \textit{IR trace analysis}.
First, LightningSim compiles the LLVM IR to an executable binary and runs it to generate an execution trace. Second, LightningSim analyzes the trace and hierarchically calculates the latency of the entire program from instructions to basic blocks and to functions.
We identify the following major challenges during the two stages:
\begin{enumerate}
    \item \textbf{Enabling execution of LLVM IR.} While the LLVM project provides the backend infrastructure capable of compiling the IR code to native machine code for popular CPU architectures~\cite{llvm}, the IR code generated by HLS is intended only for internal use and thus cannot be directly compiled into a software executable due to undefined or dummy functions. %
    We must define or redefine such functions on-the-fly to enable execution.
    \item \textbf{Enabling tracing of LLVM IR.} The scheduling information provided by the HLS tool is given in terms of the hardware start and end stages (roughly analogous to clock cycles) of each LLVM IR instruction within a function (corresponding to a hardware module). In order to accurately count clock cycles, the execution order of the LLVM IR instructions must be tracked, as well as any other events that can affect timing and stalls such as reads from and writes to FIFOs and AXI interfaces.
    \item \textbf{Dynamic schedule resolution.} HLS only provides static scheduling for each module but does not account for dynamic program behavior. During RTL generation, each \textit{stage} in the static schedule is mapped to one state of a finite state machine (FSM), which does not execute sequentially: the FSM can skip states because of branches and loops with multiple iterations. This is further complicated by pipelined loops, where the execution of stages across loop iterations can overlap. Therefore, we must recalculate the dynamic scheduling information based on the static schedule and the instruction execution trace.
    \item \textbf{Stall calculation.} Each stage of the dynamic schedule corresponds to one clock cycle except in the case of
    \textit{stalls}. Stalls can occur on attempts to read from an empty FIFO, write to a full FIFO, or read from an AXI interface that is not ready, etc.
    Therefore, we must track the state of each FIFO and AXI interface and calculate the stall cycles based on the dynamic schedule.
    \item \textbf{AXI modeling.} Stall calculation is particularly challenging for external memory accesses through AXI interfaces, which have complex internal behavior, making it difficult to accurately predict the timing of AXI transactions; FLASH~\cite{flash} does not have a timing model for this. We must design a model for AXI transactions that is as close as possible to the HLS-generated RTL design.
\end{enumerate}

In the next section, we discuss the two stages of LightningSim to address these challenges.

\section{LightningSim Techniques}
\label{sec:lighningsim}

As shown in Fig.~\ref{fig:LS-overview}, stage \textbf{trace generation} has two steps: \textbf{(A)} make LLVM IR executable; \textbf{(B)} make LLVM IR traceable, executes it, and then dumps the trace.
Stage \textbf{trace analysis} has four steps: \textbf{(C)} trace parsing; \textbf{(D)} resolving dynamic schedule; \textbf{(E)} calculating stalls; \textbf{(F)} modeling AXI interfaces.

\subsection{Making the IR Executable}
\label{sec:make-ir-executable}

HLS front-end compilation produces LLVM IR code to be used for scheduling, binding, and RTL generation. LightningSim extracts the IR code and prepares for CPU execution.

The LLVM project has a built-in code generation backend, which compiles IR code to native machine code for modern CPU architectures including x86 and x86-64~\cite{llvm}. However, as the HLS-generated LLVM IR code is not intended for CPU machine code generation, there are undefined functions and missing intrinsics that must be defined or re-implemented \textit{on-the-fly}.
The newly implemented HLS-specific intrinsics and functions are summarized in Table~\ref{tab:missing-llvm-functions}.

\begin{table}
    \centering
    \footnotesize
    \renewcommand*{\arraystretch}{0.9}
    \caption{Implemented HLS-specific functions to enable IR execution.}
    \begin{tabular}{c|c}
    \toprule
       Missing Intrinsics  & \texttt{llvm.part.select.i$N$.i$N$(...)} \\
       & \texttt{llvm.part.set.i$M$.i$M$.i$N$(...)} \\ \midrule

       Missing Functions & \texttt{\_autotb\_FifoRead\_i$N$(...)} \\
       & \texttt{\_autotb\_FifoWrite\_i$N$(...)} \\
       & \texttt{\_ssdm\_op\_ReadReq.m\_axi.p1i$N$(...)} \\
       & \texttt{\_ssdm\_op\_Read.m\_axi.p1i$N$(...)} \\
       & \texttt{\_ssdm\_op\_WriteReq.m\_axi.p1i$N$(...)} \\
       & \texttt{\_ssdm\_op\_Write.m\_axi.p1i$N$(...)} \\
       & \texttt{\_ssdm\_op\_WriteResp.m\_axi.p1i$N$(...)} \\
       & \texttt{\_ssdm\_SpecMemSelectRead(...)} \\
    \bottomrule
    \end{tabular}
    \label{tab:missing-llvm-functions}
\end{table}

\para{Implementing missing intrinsics}

Bit-wise operations, such as bit selection and manipulation, are common and efficient hardware operations.
Vitis HLS compiler frequently makes use of the bit manipulation intrinsics \texttt{llvm.part.select.*} and \texttt{llvm.part.set.*}, whose implementations are missing from the latest LLVM version.
Therefore, we re-implement these intrinsics by defining a lowering operation for each intrinsic within the LLVM \texttt{SelectionDAG}. Each lowering operation replaces the intrinsic with a sequence of bit manipulation operations that are still supported by LLVM code generators.

LightningSim uses \texttt{clang} to compile the HLS-generated IR to properly handle these intrinsics.

\para{Implementing and linking missing functions}
\label{sec:missing-functions}

\begin{figure}
    \centering
    \includegraphics[width=\linewidth]{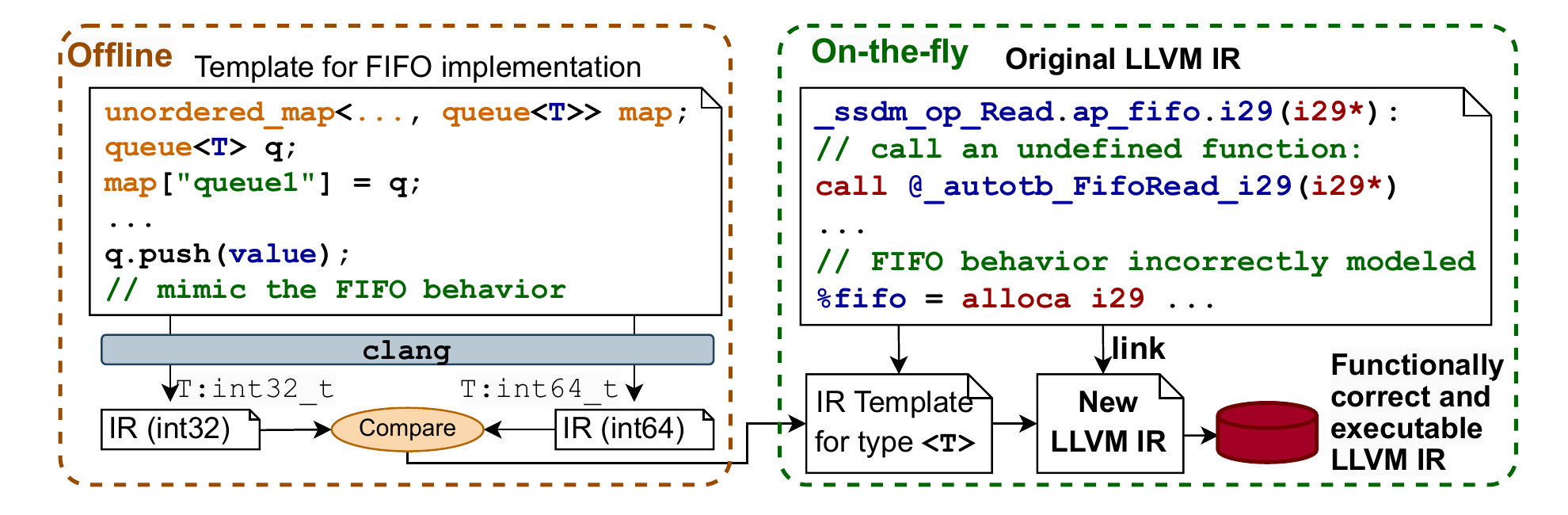}
    \caption{LightningSim implements undefined or incorrectly defined functions into new IRs on-the-fly, and then links from the original IR to build functionally correct and executable IR code.}
    \label{fig:fifo-implementation}
\end{figure}

Vitis HLS generates LLVM IR that includes references to undefined or dummy functions, which can hinder the generation of software executables and fail to accurately reflect the behavior of the HLS design. To enable IR execution, we must redefine these functions on-the-fly, as some may have arbitrary parameters that cannot be pre-defined.
A full list of LightningSim implemented functions are summarized in Table~\ref{tab:missing-llvm-functions},
where functions with names starting with \texttt{\_ssdm} or \texttt{\_autotb} are specially handled by HLS without correct implementations.

We provide one specific example for \textit{arbitrary-precision FIFO implementation}.
As shown in Fig.~\ref{fig:fifo-implementation}, the original LLVM IR 
function \texttt{\nolinkurl{\_ssdm\_op\_Read.ap\_fifo.i29(i29*)}} makes a single call to an undefined function \texttt{\nolinkurl{\_autotb\_FifoRead\_i29(i29*)}}, which is interpreted as a read from a 29-bit-wide FIFO.
To accurately calculate the design latency, during IR execution, the FIFO behaviors (en-queue, de-queue) must be faithfully reflected.
In the original LLVM IR, however, it is simply represented as a pointer to a 29-bit integer as \texttt{\%fifo = alloca i29}, which does not simulate FIFO behavior correctly.
Another challenge is that HLS allows the use of FIFOs with arbitrary-precision data types, which are not compatible with native C++.

To correctly model FIFO behavior with arbitrary-precision data types, we propose a templated C++ file, which will be compiled to LLVM IR dynamically by LightningSim. As shown in Fig.~\ref{fig:fifo-implementation}, there are two steps, offline and on-the-fly.
\underline{First}, before execution, LightningSim adopts the native C++ data structures \texttt{std::queue} to represent the FIFOs
and \texttt{std::unordered\_map} to map all FIFO pointers to actual C++ queues.
In addition, each write and read will be logged along with the FIFO pointer to a special file, which will be used by LightningSim later for trace analysis.
Furthermore, to support arbitrary-precision data types,
we create an LLVM IR template adaptive to any bitwidth $N$. We first compile two versions of the function for $N=32$ and $N=64$ to IRs, which are native C++ types \texttt{int32\_t} and \texttt{int64\_t}. By comparing the differences between the two functions, we can identify necessary changes when $N$ is an arbitrary number, and generate the correct IR function accordingly.
\underline{Second}, during execution, LightningSim reads the template, instantiates it for a given $N$, independently compiles the newly generated IR code, and links together with the original LLVM IR, and finally builds the functionally correct and executable IR code.

LightningSim uses a similar approach for other undefined or unusable functions, such as AXI read and write. The definition in the HLS-generated IR has \texttt{weak} linkage while the LightningSim definition has \texttt{external} linkage, so the LightningSim definition takes precedence during linking, thus providing the correct behavior at simulation time.

\subsection{Making the IR Traceable}
\label{sec:traceable-ir}

To create an accurate representation of hardware timing, LightningSim needs to understand the specific LLVM IR instructions that were executed and in what order during CPU simulation. Examples includes information about which conditional statements were executed and the number of iterations for each loop. Instead of identifying and handling each type of control flow construct individually within the IR, LightningSim employs a generic approach that tracks all control flow execution.

In LLVM IR and similar low-level languages, each function is composed of instructions organized in \textit{basic blocks}, which are groups of instructions with a single entry point and a single exit point. Entry points can be jump destinations of \texttt{br} (branch) and \texttt{switch}; exit points, a.k.a. \textit{terminator}, can be \texttt{br}, \texttt{switch}, or \texttt{ret} (return).

The execution of a basic block is equivalent to the sequential execution of all instructions within it. To accurately track the executed instructions with precise order, we only need to track the basic blocks that are executed. This is a popular, well-known approach to tracing CPU instructions~\cite{ball1994optimally}.

LightningSim employs a custom LLVM pass that iterates through all basic blocks and adds a call to a specific tracing function with parameters that uniquely identify each basic block. The tracing function is defined in a separate C++ file that is shipped with LightningSim and linked; it is responsible for recording the execution of each basic block, which is logged as a ``trace\_bb'' with the function name and basic block number within the function.

After the LLVM pass and the procedures outlined in Section~\ref{sec:make-ir-executable}, the IR code can be executed on CPU. The resulting trace includes trace logs from newly defined functions as described in Section~\ref{sec:missing-functions}, enabling 
the second stage of LightningSim, \textit{trace analysis}.

\subsection{Trace Parsing}
\label{sec:trace-parsing}
\begin{figure}
    \begin{center}
        \includegraphics[width=\linewidth]{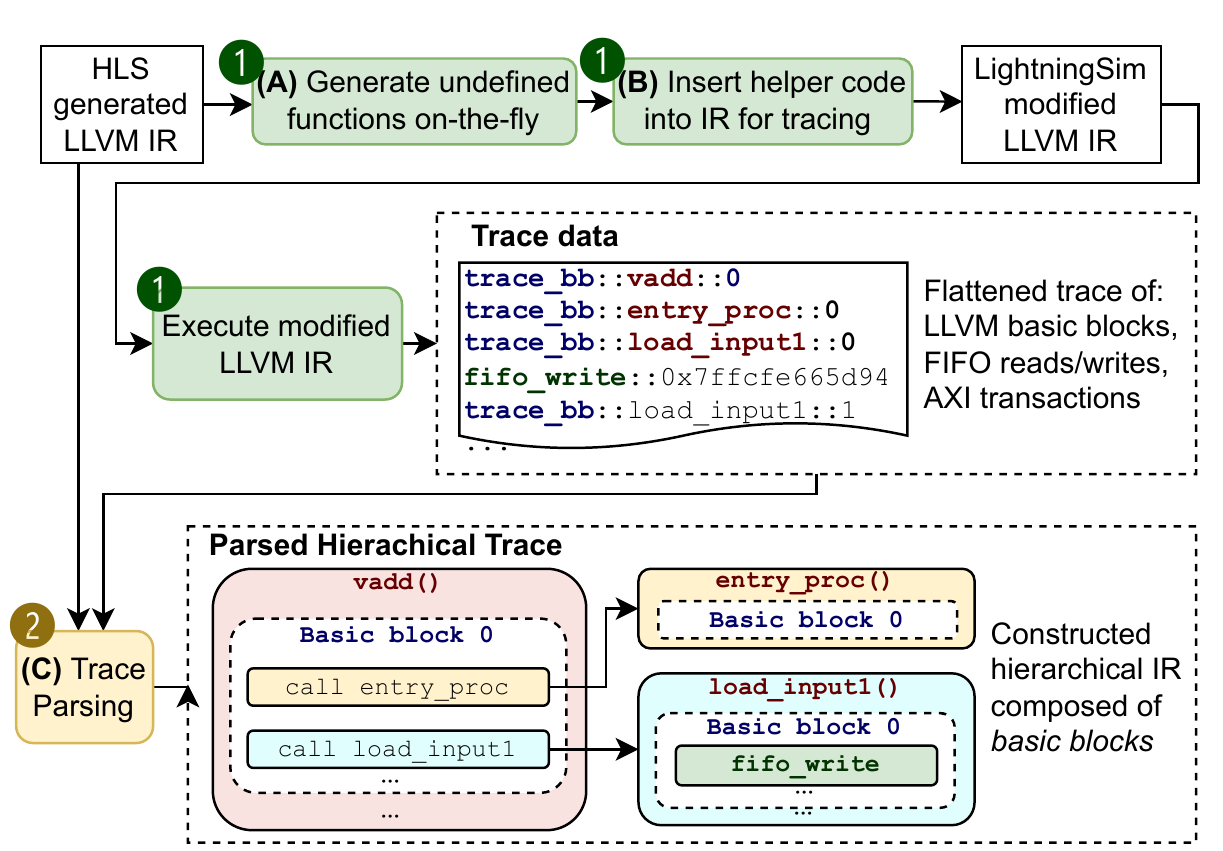}
    \end{center}
    \caption{LightningSim parses the flat execution trace into a hierarchical data structure consisting of basic blocks in each function call. Each function contains events like FIFO operations, AXI transactions, and function sub-calls. Each sub-call is recursively parsed into its own hierarchical data structure.
    }
    \label{fig:trace-parsing}
\end{figure}

Trace parsing prepares for the second stage, trace analysis.
The executed and generated trace is ``flat'', as in the example shown in Fig.~\ref{fig:trace-parsing}, the \textit{trace data}.
The flat trace data includes the exact orders of instructions, FIFOs that were read and written, and off-chip DRAM addresses that were read and written through AXI interfaces.
However, because functions in HLS become hardware modules that run in parallel, we must isolate the trace data for each function call to be processed independently, which requires trace parsing.
Therefore, LightningSim transforms the flat trace into a hierarchical structure of function calls, as shown in the bottom half of Fig.~\ref{fig:trace-parsing}. After parsing, each call consists of a list of basic blocks, each of which has a mapping of AXI, FIFO, and/or sub-call instructions to their corresponding lines in the trace. For sub-calls, the structure includes a nested list of basic blocks in a recursive format.

\subsection{Resolving the Dynamic Schedule}
\label{sec:resolving-dynamic-trace}

\begin{figure*}
    \begin{center}
        \includegraphics[width=\linewidth]{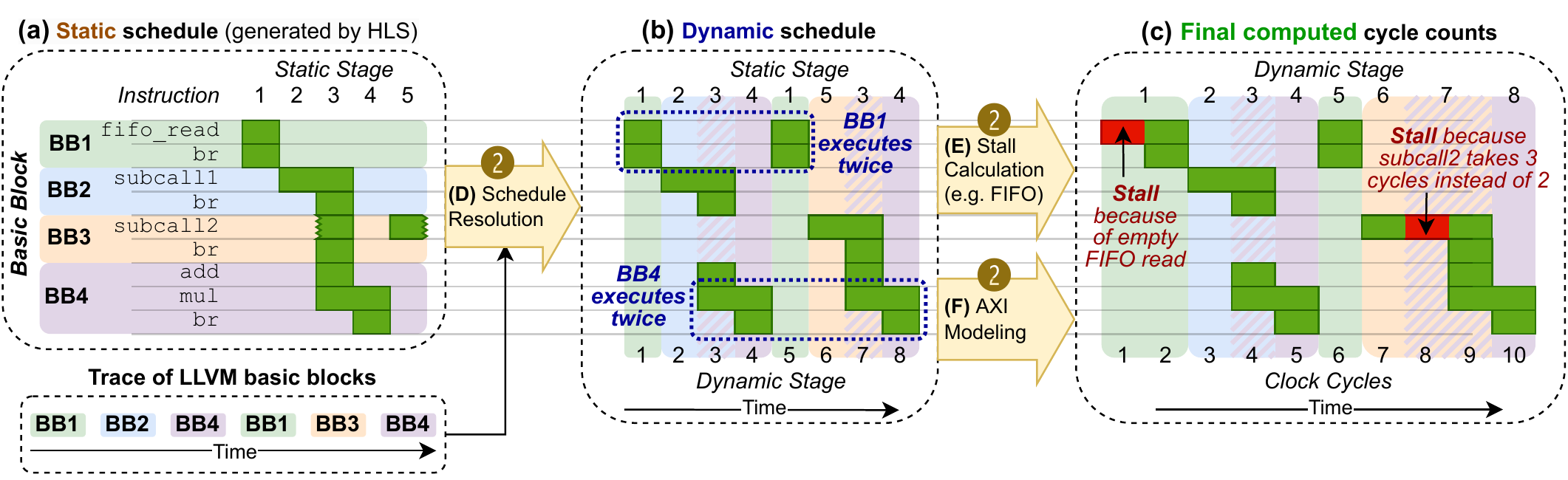}
    \end{center}
    \caption{An example of LightningSim's trace analysis process for a single function call. Using the trace of LLVM basic blocks generated during simulation (\S\ref{sec:traceable-ir}) and static schedule data from HLS synthesis, schedule resolution (\S\ref{sec:resolving-dynamic-trace}) resolves a dynamic schedule where dynamic stages increase monotonically over time. Then, stall calculation (\S\ref{sec:stall-calculation}) determines stall cycles for each stage.}
    \label{fig:trace-analysis}
\end{figure*}

Dynamic schedule resolution is a crucial step in LightningSim. As shown in Fig.~\ref{fig:trace-analysis}, it uses the \textbf{static schedule} generated by HLS and the trace generated in the first stage of LightningSim to determine the \textbf{dynamic schedule} for each instruction and function. Then, the \textbf{final clock cycle counts} are collectively computed after stall calculation and AXI modeling.

\para{Discrepancies between static and dynamic schedule}

After front-end compilation and scheduling, HLS generates a static schedule for each function, annotated by \textit{stages}, that indicates the start and end stage for each instruction. In the generated RTL, each function is mapped to a hardware module controlled by a finite state machine (FSM).
\textbf{Each stage usually corresponds to one clock cycle but with several exceptions because of the dynamic behavior} of the program, including FIFO stalls, sub-calls to other functions, and special behavior for pipelines and dataflows. 
Specifically, 
with pipeline, multiple stages can execute simultaneously; with dataflow, the starts and ends of dataflow processes (sub-calls) are determined not by their position in the schedule but by their inputs and outputs; with loops, the loop body can be executed multiple times. Therefore, to accurately calculate cycles, LightningSim needs to resolve each function's static schedule into a dynamic schedule based on the executed trace, where the stage numbers monotonically increase.

The basic unit of LightningSim's trace resolution is basic block (BB): each instance of a BB in the trace can be linked to a set of ``static stages'' in the HLS-generated schedule and a set of ``dynamic stages'' based on simulation behavior. The information from HLS helps identify which BBs belong to pipeline and dataflow regions, allowing for separate handling of three cases: non-pipelined non-dataflow BBs, pipelined BBs, and dataflow BBs.

\para{Resolving non-pipelined non-dataflow basic blocks}

\begin{algorithm}[t]
\small
  \caption{\small Resolving Dynamic Schedule%
}\label{dynamic-schedule}
  \label{alg:dynamic-schedule-resolution}
  
  \begin{algorithmic}[1]
  \Require BBs' static start/end stage, executed trace, pipeline info
  \Ensure Each BB's dynamic start/end stage

\State bb\_prev $ \leftarrow \emptyset $
\State bb\_prev.static\_start $ \leftarrow $ 0, bb\_prev.static\_end $ \leftarrow $ 0
\State bb\_prev.dynamic\_start $ \leftarrow $ 0, bb\_prev.dynamic\_end $ \leftarrow $ 0
   
\For{bb $\in$ Trace(all BBs)}

\State delay $ \leftarrow$ bb.static\_start \textminus{} bb\_prev.static\_end \label{algline:delay}

\If{non-pipeline}
    \State delay $\leftarrow$ max(delay, 1) \label{algline:clamp}
\EndIf

\If{bb initiates a new iteration}
    \If{non-pipeline}
        \State delay $\leftarrow$ 1 \Comment{current BB starts without overlap}
    \Else
        \State delay $\leftarrow$ delay + pipeline.II \Comment{overlap but add II}
    \EndIf
\EndIf

\State bb.dynamic\_start $\leftarrow$ bb\_prev.dynamic\_end + delay \label{algline:dynamic-start}

\State bb.dynamic\_end $\leftarrow$ bb.dynamic\_start + span \textminus{} 1 \label{algline:dynamic-end}

\EndFor

  \end{algorithmic}
\end{algorithm}

LightningSim handles trace resolution iteratively by updating three state variables for each basic block in the trace. Each trace resolution procedure is within a single function. The three state variables are:
\begin{itemize}
    \item The \textit{static start/end stage}: the stage in the static schedule at which the current basic block starts/ends. This is obtained from the HLS-generated schedule information.
    \item The \textit{dynamic start/end stage}: the monotonically increasing stage in the dynamic schedule at which the current basic block starts/ends. This is what LightningSim calculates for each occurrence of each basic block.
    \item The \textit{new iteration}: a flag indicating  whether a basic block initiates a new loop iteration. If it is the first in a new iteration, it begins execution immediately without delay, otherwise it adheres to the established static schedule.
    
\end{itemize}

Algorithm 1 shows the procedure of resolving the dynamic schedule. 
We use the example in Fig.~\ref{fig:trace-analysis} to explain the calculation step by step.

\noindent
\textbf{Inputs.} As shown in Fig.~\ref{fig:trace-analysis}(a), there are four BBs, from BB1 to BB4, each containing a few instructions.
Each instruction is associated with a static stage, generated by HLS scheduling. Meanwhile, there is a trace of those LLVM basic blocks, generated by LightningSim.
In this trace, BB1 and BB4 are both executed twice.
Based on this static schedule and the trace, we compute the dynamic schedule.

\noindent
\textbf{BB static start/end stage and span.}
When a new basic block is encountered in the trace, we first determine its static start and end stages. Usually, the static end stage of the BB is the static stage of the BB's terminator instruction (e.g., \texttt{br}). For example, BB1's static end stage is 1, and BB2's static end stage is 3.
Each BB is also associated with a \textit{span}, i.e., the number of stages the BB is active.
For example, BB1's span is 1, while BB2's span is 2.

We notice a special case when a BB's static start stage is not always the smallest-numbered static start stage of all instructions in the basic block. In the example in shown in Fig.~\ref{fig:trace-analysis}, basic block BB3 spans stages 3 and 5, but it actually starts at stage 5 and ends at stage 3. LightningSim identifies these special cases and uses its actual start stage.

\noindent
\textbf{BB delay.}
To decide a BB's dynamic start/end stages, we must first know how many cycles the BB has to wait after the previous BB finishes.
This value is denoted by \textit{delay}, calculated as $bb.static\_start - bb\_prev.static\_end$ (Line \ref{algline:delay}, Algorithm \ref{alg:dynamic-schedule-resolution}).
If delay is 1, it means that the current BB can start immediately after the previous BB finishes in the trace.
If delay is zero or negative, it means that the current BB can be overlapped with previous BB.
If delay is larger than 1, we always clamp it to 1 since it indicates empty stages that the FSM is able to skip in non-pipelined loops (Line \ref{algline:clamp}).

\noindent
\textbf{BB dynamic start/end stage.}
Given the previous BB's dynamic end and the current BB's delay, the current BB's dynamic start is computed as $bb\_prev.dynamic\_end + delay$ (Line \ref{algline:dynamic-start}) and dynamic end is incremented by span (Line \ref{algline:dynamic-end}).

\ding{114} \textbf{Examples} in Fig.~\ref{fig:trace-analysis}. \textbf{BB1.} As the first basic block in the trace, BB1's static and dynamic start/end stages are all 1. \textbf{BB2.} BB2 follows BB1, whose static start and end stages are 2 and 3 and span is 2.
Its delay is 1, computed as 2 (BB2's static start) minus 1 (BB1's static end).
BB2's dynamic start is 2, computed as 1 (BB1's dynamic end) plus 1 (delay), and dynamic end is 3.
\textbf{BB4.} Its delay is $3-3=0$, meaning that BB4 has one stage overlap with its previous basic block BB2, as can be seen from the static schedule. Therefore, its dynamic start is $3+0=3$ and dynamic end is 4.

\noindent
\textbf{New iteration flag.}
In non-pipelined loops, a new iteration starts as soon as the last basic block from the previous iteration finishes (we address pipelined loops later). 
To model this situation, we maintain a \textit{new iteration} flag. When this flag is set, the first BB that initiates this new iteration will begin execution one stage later than its previous BB, i.e., \textit{delay} is forced to be 1, so that there is no overlap nor empty states. Once the current BB has been updated, the flag is cleared to ensure compliance in the next iteration.

\ding{114} \textbf{Examples} in Fig.~\ref{fig:trace-analysis}. \textbf{BB1} (the second time). In this trace, BB1 is seen a second time, indicating that a new iteration starts so that BB1 can start immediately after BB4 finishes.
Thus, BB1's second dynamic start is $4+1=5$, as shown in Fig.~\ref{fig:trace-analysis}(b).
After BB1 update, the new iteration flag is cleared.
\textbf{BB3.} Next, BB3's delay is 4, computed as 5 (BB3 static start) minus 1 (BB1 static end), which should be clamped to 1 to mimic the FSM behavior of skipping empty stages. Therefore, its dynamic start is $5+1=6$ and end is 7.
\textbf{BB4.} Since BB4 is still within the second iteration, it adheres to the static schedule. Therefore, its dynamic start is $7+0=7$.
As seen at the bottom of Fig.~\ref{fig:trace-analysis}(b), the dynamic stages are from 1 to 8.

\para{Resolving pipelined basic blocks}

In pipelined loops, each loop iteration is delayed by the \textit{initiation interval} (II), which is the number of clock cycles (dynamic stages) between the beginning of one iteration and the next. For instance, if a loop's II is 2, the first iteration would start at dynamic stage 0, the second at dynamic stage 2, the third at dynamic stage 4, and so on.
Therefore, processing basic blocks within a pipelined loop has the following key distinctions compared with non-pipelined loops.

\underline{First}, 
the \textit{delay} calculation is different.
LightningSim does not clamp the delay value for basic blocks within a pipeline, as it represents a conditional statement that is not executed. In non-pipelined loops, the FSM can bypass stages that are not executed, however in pipelined loops, each iteration must have a fixed number of stages and thus no stages can be skipped.
\underline{Second}, when a new iteration starts according to the \textit{new iteration} flag, the delay value will be the loop II, which aligns with the pipelined loop behavior.
\underline{Third}, when LightningSim encounters a basic block outside the pipelined loop, it resets its \textit{static and dynamic stages} to be the maximum static and dynamic stages encountered inside the loop, ensuring that the pipelined stages do not overlap with non-pipelined stages.

\para{Resolving dataflow basic blocks}

The static start and end stages of instructions within a dataflow region do not accurately reflect their behavior in hardware. Instead, the start and end of dataflow processes are driven by their inputs and outputs. Therefore, LightningSim recalculates static start and end stages for instructions within a dataflow region based on their input and output channels and dataflow processes.

For the start stage, the rules are: (1) If the dataflow process has no inputs, it starts immediately, at stage 0. (2) If the dataflow process has any scalar inputs---i.e., inputs that are single values rather than channels---it starts one stage after all scalar inputs are produced. (3) Otherwise, the start of the dataflow process is determined by start propagation: the process starts one stage after any of its input processes start.

For the end stage, the rules are: (1) If the dataflow process has any outputs, and they are all scalar outputs (single values, not channels), then it ends in the same stage it starts. (2) Otherwise, the process synchronizes with every other process at the end of the dataflow region---its end stage is the same as all other non-scalar-output dataflow processes (determined as the latest start stage of any of them).

By using this model to recalculate the static start and end stages of every instruction in the dataflow region, we can reuse the same resolution process for dataflows as for non-pipelined non-dataflow regions in Algorithm \ref{alg:dynamic-schedule-resolution}.

\subsection{Calculating Stalls}
\label{sec:stall-calculation}

The dynamic schedule produces a dynamic stage for each instruction. However, two issues remain in counting cycles precisely. First, some stages may take more than one cycle to complete due to stalls, for instance, if a FIFO or AXI interface is not ready by the time the HLS design expects it to be. 
In addition, a too-small FIFO depth may cause deadlock, which must be detected during the simulation.
Second, within any function, stalls may delay the start of any sub-calls, such that the timing of a callee depends on its caller.
Therefore, the stalls must be calculated hierarchically and globally: the stall of a function may need to be propagated to other functions and its own caller/callee.

\noindent
\textbf{Stall simulator variables.}
LightningSim is an event-based simulator tracking every function's start and stop cycles.
It creates a set of \textit{active simulators} by initiating a new \textit{simulator} for the top-level HLS kernel function first and followed by simulators for other functions when encountered in the trace. 
Each simulator maintains its variables including:
\begin{itemize}
    \item \textit{Start cycle} and \textit{current cycle}, both initialized to zero for the top-level function. At any time, a caller's current cycle will be its callees' start cycle.

    \item \textit{Stall event}: the simulator encounters a stall, which can be a sub-call to another function or an I/O operation on a FIFO or AXI bus.

    \item \textit{Stall release}: the stall will be released if the sub-call finishes or the FIFO/AXI is ready.
\end{itemize}

\label{par:simulation-events}%
Each simulator takes as input the resolved dynamic trace from the previous step and generates a list of \textit{simulation events} within this function that might affect other functions. Each event is mapped to a dynamic stage. This list of \textit{simulation events} includes the following:
\begin{itemize}
    \item \textit{Sub-call start:} the dynamic stage at which a sub-call starts executing. In RTL, this corresponds to the FSM stage of a caller when it first asserts the callee's \texttt{ap\_start} signal.
    
    \item \textit{Sub-call end:} the last dynamic stage spanned by a sub-call. In hardware, the FSM will not advance past this stage until the callee asserts its \texttt{ap\_done} signal.
    
    \item \textit{I/O event:} the dynamic stage at which an I/O operation on a FIFO or AXI bus occurs. The hardware FSM will not advance past this stage until the I/O operation succeeds.

\end{itemize}

\noindent
\textbf{Simulator workflow.}
Starting from the top-level function, LightningSim proceeds to step through the list of the simulation events for each active simulator (initially only the top-level simulator). Within a simulator, for every dynamic stage, its current cycle is incremented by one. If a sub-call start event is encountered, a new simulator for the callee function will be created if does not exist, and the start cycle of the callee's simulator is the current cycle of the caller.
This new simulator is also added to LightningSim's set of active simulators.

When traversing each active simulator's events, if a stall event is encountered, LightningSim must decide its stall release cycle. If the stall event is a sub-call, the stall can be released when the callee is finished; if the stall event is a FIFO/AXI operation, the stall can be released when the I/O is ready.
LightningSim maintains data structures for every FIFO and AXI interface to precisely model the I/O status.
Once a stall event is finished, LightningSim updates the current cycle values for all simulators that were blocked by the event, as well as the data structures for FIFOs/AXI interfaces.
This process repeats until all simulators are done. By the end of this process, LightningSim has calculated the start and end cycles of every function called in the design.

\ding{114} \textbf{Examples} in Fig.~\ref{fig:trace-analysis}. We demonstrate two stall events in this example: one for BB1 because of a FIFO read event, and one for BB3 because of a sub-call event.
BB1 is assumed to be stalled for one cycle because of FIFO read, and BB3 is assumed to be stalled for one cycle because it encounters a sub-call to another function, which takes 3 cycles.
Finally, while the dynamic stages are from 1 to 8, the precise clock cycles are from 1 to 10 because of these stalls.

\noindent
\textbf{Deadlock detection.}
LightningSim is able to detect any deadlocks during simulation using a method inspired by works on buffer lock~\cite{parks1995bounded}. If LightningSim cannot calculate the stall release cycle for any of the stall events, it indicates the hardware design will deadlock. This can occur due to cyclic dependencies in FIFO I/O: for example, module $A$ is blocked on a write to a full FIFO $X$ that would later be read by module $B$; module $B$ is blocked on a read from an empty FIFO $Y$, which is expected to be written by module $A$. In this scenario, LightningSim will identify that none of the simulators can proceed, i.e., a deadlock.

\noindent
\textbf{Incremental simulation.}
Notably, once LightningSim detects a deadlock, it is able quickly to suggest solutions by changing the FIFO depth and incrementally re-run the step of \textit{calculating stalls} only, while keeping the executed trace and dynamically resolved schedule unchanged.
Since calculating stalls is very efficient (within seconds), LightningSim can quickly identify an \textbf{optimal} FIFO depth %
by calculating the maximum depth observed when the simulation depth is unlimited.

This is fundamentally different from both HLS tools and existing simulators: once the FIFO depth is changed, they require a full-blown re-synthesis and re-run of the entire RTL generation and simulation, while LightningSim does not, benefiting from our \textbf{decoupled} two stages.

\subsection{Modeling AXI Interfaces}

The HLS-generated design incorporates its own AXI controllers for each AXI interface, which mediates requests and responses on all AXI buses between the HLS design and the rest of the system. These controllers are used, for example, to break long AXI bursts generated internally from the HLS design into multiple specification-compliant AXI bursts that do not cross 4~KB address boundaries. As the AXI controllers are implemented only in the generated RTL, LightningSim uses a detailed model to calculate the stall cycles for AXI I/O events in a manner as close as possible to the complex RTL-described behavior of the AXI controllers.

For each AXI read or write request issued within the HLS design, LightningSim first calculates the request's \textit{burst count}, i.e., how many bursts are required for this request in case it crosses the 4~KB address boundaries. To model the behavior of the \texttt{fifo\_rctl}, which holds up to 16 outstanding bursts at a time, LightningSim tracks \textit{rctl depth}, which is incremented for every outstanding request by the burst count and decremented by the same amount once all read/write data is consumed for that request. Any requests that would increase \textit{rctl depth} past 16 outstanding bursts are enqueued onto a \textit{pending} queue and issued once \textit{rctl depth} has decreased to accommodate them.

Overheads for AXI reads and write responses are modeled by a fixed number of cycles, determined empirically, atop the AXI interface latency specified by the HLS design using \texttt{\#pragma HLS interface}. This closely models the behavior of Vitis HLS C/RTL co-simulation, which incurs a fixed number of cycles as overhead on each AXI transaction due to internal logic in the HLS-generated AXI controller and the SystemVerilog AXI testbench.

\section{Execution Time}

LightningSim's time complexity can be represented as roughly \((\)time to compile and run LLVM IR\()\) + \(O(\)\#~BBs in trace\()\) + \(O(\)\#~function calls in trace\()\) + \(O(\)\#~FIFO/AXI reads/writes in trace\()\), derived from the sum of the following:
\begin{enumerate}
    \item \textbf{Trace generation.} This is entirely dependent on how long the compilation and execution of the LLVM IR takes.
    \item \textbf{Trace parsing.} This is linear with respect to the number of entries in the trace. The trace contains one entry for each of these events:
    \begin{itemize}
        \item Execution of a basic block
        \item A read or write of a FIFO
        \item An AXI bus operation (\texttt{read}, \texttt{write}, \texttt{readreq}, \texttt{writereq}, or \texttt{writeresp})
    \end{itemize}
    \item \textbf{Schedule resolution.} This is also linear with respect to the number of trace events, as described above.
    \item \textbf{Stall calculation.} This is roughly linear with respect to the number of \textit{simulation events} (see Sec.~\ref{par:simulation-events}) in the trace. This includes:
    \begin{itemize}
        \item Start of a sub-call (anywhere in the function call graph)
        \item End of a sub-call (anywhere in the function call graph)
        \item A read or write of a FIFO
        \item An AXI bus operation (\texttt{read}, \texttt{write}, \texttt{readreq}, \texttt{writereq}, or \texttt{writeresp})
    \end{itemize}
\end{enumerate}
When changing FIFO depths, only the last step needs to be recomputed, leading to a time complexity of about \(O(\)\#~function calls in trace\()\) + \(O(\)\#~FIFO/AXI reads/writes in trace\()\).

\section{LightningSim Usage}

\begin{figure}
    \begin{center}
        \includegraphics[width=\linewidth]{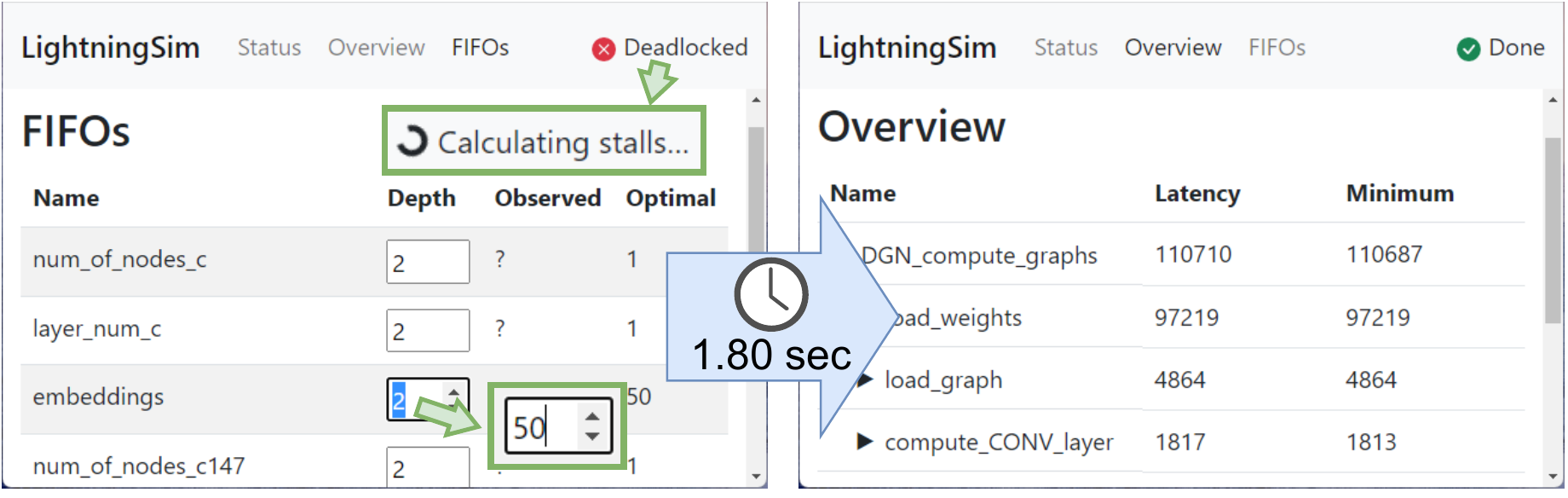}
    \end{center}
    \caption{The LightningSim user interface. FIFO depths can be changed on-the-fly. This example shows the UI for FlowGNN DGN~\cite{flowgnn}, which takes 1.80 seconds for stall recalculation.}
    \label{fig:gui}
\end{figure}

We now describe the LightningSim usage workflow. LightningSim is open-source and can be downloaded and installed following the instructions in our repository.\footnoteref{footnote:github}

LightningSim is designed for push-button ease of use. It is shipped as a CLI Python script which is invoked as follows: \texttt{./main.py \nolinkurl{/path/to/Vitis/HLS/project/solution1}}. This starts a web server, which hosts a web UI allowing the user to view the progress of the LightningSim flow and control its execution, as shown in Fig.~\ref{fig:gui}.

LightningSim reads all project settings directly from the Vitis HLS solution directory, including testbench files, compilation flags, etc.\ so as to match the C simulation settings of Vitis HLS as closely as possible. This means that \textbf{no additional configuration is required} for LightningSim to work with an existing Vitis HLS project.

By default, LightningSim will wait for the next HLS synthesis run to start before beginning its simulation, though this can be skipped if the user wants to run simulation using an already-completed HLS synthesis. At each step, LightningSim waits for only the necessary HLS-generated files, allowing it to start LLVM instrumentation as soon as the architecture synthesis phase of HLS has been completed, and to start trace analysis as soon as the HLS schedule data is generated.

Once LightningSim simulation is complete, the \textit{Overview} tab of the web UI provides a hierarchical tree view of the latency in clock cycles of each function call, along with a \textit{minimum} latency for each call. The minimum latency is calculated by assuming all FIFOs have infinite depth and determining the resulting latency, which represents the minimum possible latency that can be obtained by adjusting FIFO depths. LightningSim's trace-based analysis approach enables this minimum to be calculated efficiently by simply re-running stall calculation with different parameters.

The web UI also exposes a \textit{FIFOs} tab containing a table with information about each FIFO used in the design. Each FIFO is listed with a name, depth, observed depth, and optimal depth. The observed depth represents the maximum queue length seen for that FIFO at any clock cycle during the simulation; the optimal depth is the same measurement for a simulation variant where all FIFOs are unbounded, as is used to calculate the \textit{minimum latencies} previously described.

Editable numeric fields allows the user to adjust the depth of each FIFO in the simulation. Upon adjusting the depth, LightningSim automatically recalculates the latency and observed FIFO depths in the background \textbf{without re-running the entire simulation.} This allows the user to quickly explore the impact of FIFO depth on the speed of their hardware design.

\section{Results}

\subsection{Evaluation Setup}

We evaluate LightningSim on a server running Red Hat Enterprise Linux 7.9 using version 2021.1 of the Xilinx tools, including Xilinx Vitis HLS 2021.1. The server has a 64-core Intel Xeon Gold 6226R CPU and 502~GiB of RAM. For each test case, we compare the cycle counts and runtime of LightningSim against those of synthesis using Vitis HLS and C/RTL cosimulation using Vivado XSIM.

\subsection{Benchmarks}

We use 33 benchmarks to comprehensively evaluate LightningSim's accuracy and performance.
As shown in Table~\ref{tab:results} second column, each benchmark has up to five advanced features to evaluate LightningSim, including: \textbf{C} for sub-calls to other functions; \textbf{P} for pipelined loops; \textbf{D} for dataflow regions; \textbf{F} for FIFO streams; \textbf{A} for AXI master interfaces.
Part of the benchmarks come from Xilinx sample code repositories~\cite{xilinx2021basic,xilinx2022vitis}, containing many small benchmarks that test a wide variety of HLS features. We also use several examples of Vitis HLS projects from Kastner \textit{et al.}'s book \textit{Parallel Programming for FPGAs}~\cite{kastner2018parallel}, which includes implementations of several popular algorithms in HLS. Finally, for a comprehensive test, we use five designs from FlowGNN~\cite{flowgnn}, the most recent and open-sourced HLS designs of five graph neural networks (GNNs) accelerators extensively using dataflow, pipelines, data streaming, and AXI interfaces.
Each GNN implementation in FlowGNN has over 1500 lines of code and 23 functions. 

\subsection{Performance and Accuracy}

\begin{table*}
    \caption{Speed and accuracy metrics of LightningSim on various benchmarks.}
    \label{tab:results}
    \centering
    \setlength{\tabcolsep}{2.5pt}
    \renewcommand*{\arraystretch}{0.925}
    \footnotesize
    \begin{tabular}{l|ccccc|ccccc|ccc}
        \toprule
        & \multicolumn{5}{c|}{\textbf{Features}} & \multicolumn{5}{c|}{\textbf{Time (s)}} & \multicolumn{3}{c}{\textbf{Cycles}} \\
        \textbf{Benchmark} & \textbf{C} & \textbf{P} & \textbf{D} & \textbf{F} & \textbf{A} & \textbf{HLS} & \textbf{LS\,$\mathbf{\parallel}$\,HLS} & \textbf{Cosim} & \textbf{LS} & \textbf{LS Inc} & \textbf{Cosim} & \textbf{LS} & \textbf{HLS} \\
        \midrule
        Fixed-point square root~\cite{xilinx2021basic} & \xmark & \cmark & \xmark & \xmark & \xmark & 9.83 & 13.90 {\scriptsize (+41.5\%)} & 27.25 & 4.87 {\scriptsize (5.6\texttimes)} & 0.00 & 30 & 30 {\scriptsize (\textpm 0.0\%)} & 32 {\scriptsize (+6.7\%)} \\
        FIR filter~\cite{xilinx2021basic} & \xmark & \cmark & \xmark & \xmark & \xmark & 7.60 & 8.52 {\scriptsize (+12.0\%)} & 20.12 & 1.76 {\scriptsize (11.5\texttimes)} & 0.00 & 172 & 172 {\scriptsize (\textpm 0.0\%)} & 174 {\scriptsize (+1.2\%)} \\
        Fixed-point window conv~\cite{xilinx2021basic} & \xmark & \cmark & \xmark & \xmark & \xmark & 12.00 & 14.20 {\scriptsize (+18.3\%)} & 28.30 & 3.14 {\scriptsize (9.0\texttimes)} & 0.00 & 35 & 35 {\scriptsize (\textpm 0.0\%)} & 37 {\scriptsize (+5.7\%)} \\
        Floating point conv~\cite{xilinx2021basic} & \xmark & \cmark & \xmark & \xmark & \xmark & 8.34 & 9.29 {\scriptsize (+11.5\%)} & 49.78 & 1.80 {\scriptsize (27.6\texttimes)} & 0.00 & 35 & 35 {\scriptsize (\textpm 0.0\%)} & 37 {\scriptsize (+5.7\%)} \\
        Arbitrary precision ALU~\cite{xilinx2021basic} & \xmark & \xmark & \xmark & \xmark & \xmark & 9.37 & 10.18 {\scriptsize (+8.5\%)} & 24.17 & 1.59 {\scriptsize (15.2\texttimes)} & 0.00 & 36 & 36 {\scriptsize (\textpm 0.0\%)} & 36 {\scriptsize (\textpm 0.0\%)} \\
        Parallel loops~\cite{xilinx2021basic} & \cmark & \cmark & \xmark & \xmark & \xmark & 10.97 & 11.83 {\scriptsize (+7.9\%)} & 26.81 & 1.70 {\scriptsize (15.7\texttimes)} & 0.00 & 32 & 32 {\scriptsize (\textpm 0.0\%)} & 34 {\scriptsize (+6.2\%)} \\
        Imperfect loops~\cite{xilinx2021basic} & \cmark & \cmark & \xmark & \xmark & \xmark & 11.95 & 11.94 {\scriptsize (\textminus 0.1\%)} & 25.80 & 1.66 {\scriptsize (15.5\texttimes)} & 0.00 & 34 & 34 {\scriptsize (\textpm 0.0\%)} & 36 {\scriptsize (+5.9\%)} \\
        Loop with max bound~\cite{xilinx2021basic} & \xmark & \cmark & \xmark & \xmark & \xmark & 10.86 & 12.04 {\scriptsize (+10.9\%)} & 24.76 & 1.71 {\scriptsize (14.4\texttimes)} & 0.00 & 31 & 31 {\scriptsize (\textpm 0.0\%)} & 33 {\scriptsize (+6.5\%)} \\
        Perfect nested loops~\cite{xilinx2021basic} & \xmark & \cmark & \xmark & \xmark & \xmark & 11.48 & 12.30 {\scriptsize (+7.1\%)} & 24.76 & 1.70 {\scriptsize (14.6\texttimes)} & 0.00 & 406 & 406 {\scriptsize (\textpm 0.0\%)} & 408 {\scriptsize (+0.5\%)} \\
        Pipelined nested loops~\cite{xilinx2021basic} & \xmark & \cmark & \xmark & \xmark & \xmark & 11.34 & 12.42 {\scriptsize (+9.5\%)} & 24.92 & 1.67 {\scriptsize (14.9\texttimes)} & 0.00 & 405 & 405 {\scriptsize (\textpm 0.0\%)} & 405 {\scriptsize (\textpm 0.0\%)} \\
        Sequential accumulators~\cite{xilinx2021basic} & \cmark & \cmark & \xmark & \xmark & \xmark & 11.46 & 12.14 {\scriptsize (+5.9\%)} & 26.59 & 1.76 {\scriptsize (15.1\texttimes)} & 0.00 & 32 & 32 {\scriptsize (\textpm 0.0\%)} & 34 {\scriptsize (+6.2\%)} \\
        Accumulators + asserts~\cite{xilinx2021basic} & \cmark & \cmark & \xmark & \xmark & \xmark & 11.62 & 12.39 {\scriptsize (+6.6\%)} & 27.13 & 1.74 {\scriptsize (15.5\texttimes)} & 0.00 & 33 & 33 {\scriptsize (\textpm 0.0\%)} & 259 {\scriptsize (+684.8\%)} \\
        Accumulators + dataflow~\cite{xilinx2021basic} & \cmark & \cmark & \cmark & \xmark & \xmark & 11.99 & 12.50 {\scriptsize (+4.3\%)} & 27.26 & 1.79 {\scriptsize (15.2\texttimes)} & 0.00 & 31 & 31 {\scriptsize (\textpm 0.0\%)} & 33 {\scriptsize (+6.5\%)} \\
        Static memory example~\cite{xilinx2021basic} & \cmark & \cmark & \xmark & \xmark & \xmark & 6.18 & 6.92 {\scriptsize (+12.0\%)} & 33.23 & 1.65 {\scriptsize (20.2\texttimes)} & 0.00 & 66 & 66 {\scriptsize (\textpm 0.0\%)} & 70 {\scriptsize (+6.1\%)} \\
        Pointer casting example~\cite{xilinx2021basic} & \xmark & \cmark & \xmark & \xmark & \xmark & 5.01 & 6.03 {\scriptsize (+20.4\%)} & 32.55 & 1.58 {\scriptsize (20.6\texttimes)} & 0.00 & 408 & 408 {\scriptsize (\textpm 0.0\%)} & 410 {\scriptsize (+0.5\%)} \\
        Double pointer example~\cite{xilinx2021basic} & \cmark & \cmark & \xmark & \xmark & \xmark & 5.57 & 6.17 {\scriptsize (+10.7\%)} & 31.70 & 1.47 {\scriptsize (21.5\texttimes)} & 0.00 & 25 & 25 {\scriptsize (\textpm 0.0\%)} & 29 {\scriptsize (+16.0\%)} \\
        AXI4 master~\cite{xilinx2021basic} & \cmark & \cmark & \xmark & \xmark & \cmark & 6.36 & 6.63 {\scriptsize (+4.3\%)} & 21.06 & 1.64 {\scriptsize (12.9\texttimes)} & 0.00 & 178 & 177 {\scriptsize (\textminus 0.6\%)} & 176 {\scriptsize (\textminus 1.1\%)} \\
        AXIS w/o side channel~\cite{xilinx2021basic} & \xmark & \cmark & \xmark & \xmark & \xmark & 5.29 & 6.34 {\scriptsize (+19.8\%)} & 19.12 & 1.51 {\scriptsize (12.7\texttimes)} & 0.00 & 52 & 51 {\scriptsize (\textminus 1.9\%)} & 53 {\scriptsize (+1.9\%)} \\
        Multiple array access~\cite{xilinx2021basic} & \xmark & \cmark & \xmark & \xmark & \xmark & 11.05 & 12.03 {\scriptsize (+8.8\%)} & 24.32 & 1.59 {\scriptsize (15.3\texttimes)} & 0.00 & 252 & 252 {\scriptsize (\textpm 0.0\%)} & 254 {\scriptsize (+0.8\%)} \\
        Resolved array access~\cite{xilinx2021basic} & \cmark & \cmark & \xmark & \xmark & \xmark & 10.55 & 11.55 {\scriptsize (+9.5\%)} & 24.36 & 1.61 {\scriptsize (15.2\texttimes)} & 0.00 & 131 & 131 {\scriptsize (\textpm 0.0\%)} & 133 {\scriptsize (+1.5\%)} \\
        URAM with ECC~\cite{xilinx2021basic} & \cmark & \cmark & \xmark & \xmark & \xmark & 7.32 & 6.38 {\scriptsize (\textminus 12.9\%)} & 22.07 & 1.71 {\scriptsize (12.9\texttimes)} & 0.00 & 115 & 115 {\scriptsize (\textpm 0.0\%)} & 121 {\scriptsize (+5.2\%)} \\
        Fixed-point Hamming~\cite{xilinx2021basic} & \xmark & \cmark & \xmark & \xmark & \xmark & 8.71 & 9.73 {\scriptsize (+11.7\%)} & 33.28 & 1.87 {\scriptsize (17.8\texttimes)} & 0.00 & 259 & 259 {\scriptsize (\textpm 0.0\%)} & 261 {\scriptsize (+0.8\%)} \\
        Unoptimized FFT~\cite{kastner2018parallel} & \cmark & \cmark & \xmark & \xmark & \xmark & 15.90 & 14.89 {\scriptsize (\textminus 6.3\%)} & 153.53 & 4.28 {\scriptsize (35.9\texttimes)} & 1.85 & 261,781 & 261,150 {\scriptsize (\textminus 0.2\%)} & ? \\
        Multi-stage FFT~\cite{kastner2018parallel} & \cmark & \cmark & \cmark & \xmark & \xmark & 17.33 & 12.95 {\scriptsize (\textminus 25.3\%)} & 61.43 & 2.07 {\scriptsize (29.7\texttimes)} & 0.00 & 3,770 & 3,772 {\scriptsize (+0.1\%)} & ? \\
        Huffman encoding~\cite{kastner2018parallel} & \cmark & \cmark & \cmark & \xmark & \xmark & 43.82 & 38.79 {\scriptsize (\textminus 11.5\%)} & 46.89 & 2.29 {\scriptsize (20.5\texttimes)} & 0.01 & 10,283 & 10,272 {\scriptsize (\textminus 0.1\%)} & ? \\
        Matrix multiplication~\cite{kastner2018parallel} & \xmark & \cmark & \xmark & \xmark & \xmark & 6.87 & 6.81 {\scriptsize (\textminus 0.8\%)} & 26.33 & 1.84 {\scriptsize (14.3\texttimes)} & 0.00 & 1,036 & 1,036 {\scriptsize (\textpm 0.0\%)} & 1,038 {\scriptsize (+0.2\%)} \\
        Parallelized merge sort~\cite{kastner2018parallel} & \cmark & \cmark & \cmark & \xmark & \xmark & 8.04 & 6.30 {\scriptsize (\textminus 21.7\%)} & 48.79 & 1.59 {\scriptsize (30.7\texttimes)} & 0.00 & 131 & 131 {\scriptsize (\textpm 0.0\%)} & 139 {\scriptsize (+6.1\%)} \\
        Vector add with stream~\cite{xilinx2022vitis} & \cmark & \cmark & \cmark & \cmark & \cmark & 11.04 & 10.01 {\scriptsize (\textminus 9.3\%)} & 27.21 & 3.74 {\scriptsize (7.3\texttimes)} & 0.36 & 4,261 & 4,261 {\scriptsize (\textpm 0.0\%)} & 4,242 {\scriptsize (\textminus 0.4\%)} \\
        FlowGNN GIN~\cite{flowgnn} & \cmark & \cmark & \cmark & \cmark & \cmark & 824.12 & 566.17 {\scriptsize (\textminus 31.3\%)} & 4219.85 & 44.01 {\scriptsize (95.9\texttimes)} & 2.62 & 260,359 & 260,337 {\scriptsize (\textminus 0.0\%)} & 216,007 {\scriptsize (\textminus 17.0\%)} \\
        FlowGNN GCN~\cite{flowgnn} & \cmark & \cmark & \cmark & \cmark & \cmark & 3224.02 & 3022.28 {\scriptsize (\textminus 6.3\%)} & 534.33 & 53.65 {\scriptsize (10.0\texttimes)} & 19.67 & 112,836 & 112,561 {\scriptsize (\textminus 0.2\%)} & 68,170 {\scriptsize (\textminus 39.6\%)} \\
        FlowGNN GAT~\cite{flowgnn} & \cmark & \cmark & \cmark & \cmark & \cmark & 965.50 & 723.50 {\scriptsize (\textminus 25.1\%)} & 838.24 & 42.87 {\scriptsize (19.6\texttimes)} & 14.15 &  17,282 & 17,282 {\scriptsize (\textpm 0.0\%)} & 16,102 {\scriptsize (\textminus 6.8\%)} \\
        FlowGNN PNA~\cite{flowgnn} & \cmark & \cmark & \cmark & \cmark & \cmark & 820.31 & 592.29 {\scriptsize (\textminus 27.8\%)} & 3285.45 & 56.02 {\scriptsize (58.7\texttimes)} & 8.54 & 344,206 & 344,206 {\scriptsize (\textpm 0.0\%)} & ? \\
        FlowGNN DGN~\cite{flowgnn} & \cmark & \cmark & \cmark & \cmark & \cmark & 848.94 & 606.15 {\scriptsize (\textminus 28.6\%)} & 996.13 & 38.46 {\scriptsize (25.9\texttimes)} & 1.80 & 110,710 & 110,710 {\scriptsize (\textpm 0.0\%)} & 93,846 {\scriptsize (\textminus 15.2\%)} \\
        \bottomrule
    \end{tabular}

    \vspace{0.5em}

    \begin{tabular}{l}
        \footnotesize
        \textbf{Features} indicate whether benchmark has: \textbf{C}: sub-\textbf{c}alls to other functions; \textbf{P:} \textbf{p}ipelined loops; \textbf{D:} \textbf{d}ataflow regions; 
        \textbf{F:} \textbf{F}IFO streams;      \textbf{A:} \textbf{A}XI interfaces. \\
        \textbf{HLS:} Time for HLS synthesis.
        \textbf{LS\,$\mathbf{\parallel}$\,HLS:}  Time from HLS synthesis starts to LightningSim finishes in parallel (vs.\ HLS synthesis alone). \\
        \textbf{Cosim:} Time for C/RTL co-simulation. \textbf{LS:} Time for LightningSim end-to-end simulation after HLS synthesis is completed (vs.\ C/RTL co-simulation). \\
        \textbf{LS Inc.:} Time for LightningSim to incrementally re-calculate stalls after changing FIFO depths. \\
        \textbf{Cosim Cycles:} Clock cycle counts reported by C/RTL co-simulation.        \textbf{LS Cycles:} Clock cycles counts reported by LightningSim (vs.\ C/RTL co-simulation). \\
        \textbf{HLS Cycles:} Clock cycles counts reported by HLS synthesis report, when available (vs.\ C/RTL co-simulation). \\
    \end{tabular}
\end{table*}

\begin{figure}
    \begin{center}
        \includegraphics[width=\linewidth]{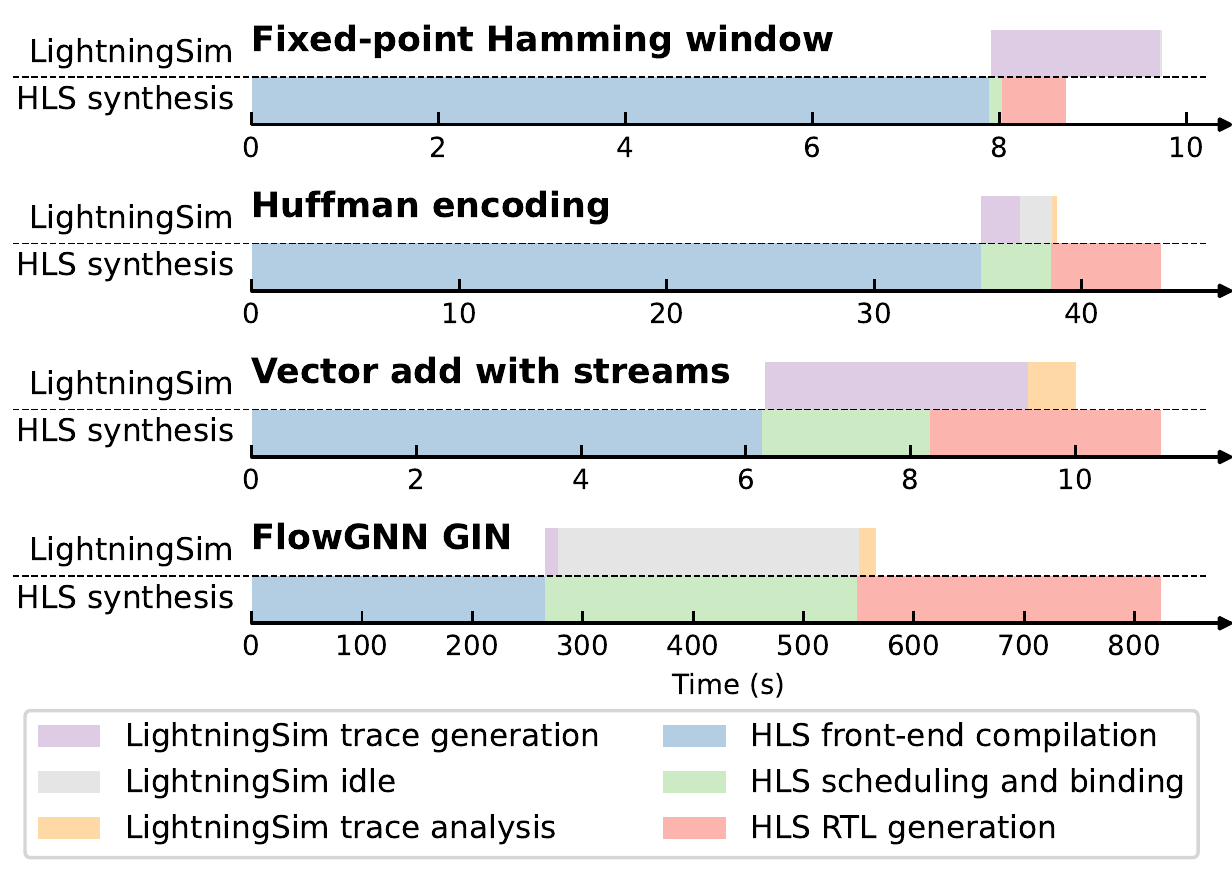}
    \end{center}
    \caption{Timelines of the parallel execution of LightningSim and HLS synthesis on various benchmarks. LightningSim's parallelism is particularly advantageous for complex designs where HLS scheduling/binding and RTL generation take minutes.}
    \label{fig:timelines}
\end{figure}

Performance and accuracy results are shown in Table~\ref{tab:results}. We note that we do not distinguish between simulation preparation and execution: we report \textit{end-to-end} timing for LightningSim.
The table shows two categories, \textbf{Time} in terms of seconds, and end-to-end \textbf{Cycles}.

\noindent
\textbf{Time.}
First, comparing with C/RTL cosimulation, LightningSim achieves between 5.6\texttimes{} and 95.9\texttimes{} speedup  (\textbf{LS} vs. \textbf{Cosim}). For larger designs such as FlowGNN models, the speedup is usually more significant. This comparison excludes HLS synthesis time, i.e., before LightningSim and cosim, we assume that HLS synthesis is finished.

Second, when LightningSim and HLS synthesis execute in parallel, LightningSim finishes faster than HLS synthesis in 13 cases out of 33, shown as negative values in the column \textbf{LS\,$\parallel$\,HLS}.
For larger designs such as FlowGNN, LightningSim always finishes faster than HLS synthesis.
In all cases where HLS synthesis completes before LightningSim, both synthesis and LightningSim finish within 15 seconds of synthesis start. 

Third, LightningSim achieves its largest speedups in incremental stall recalculation (\textbf{LS Inc}). 
This step changes the FIFO depth and re-executes the simulation to compute stalls and clock cycles.
This process takes under 20 seconds in all test cases, enabling hardware designers to evaluate deadlocks and stalls in streaming dataflow designs with unprecedented speed.

Unfortunately, neither FLASH~\cite{flash} nor FastSim~\cite{fastsim} have open-sourced their tools at the time of writing, making it difficult to compare. However, we note that FLASH's simulation preparation + simulation time averages 58\% slower than HLS synthesis alone, which is significantly slower than LightningSim's average of 1.04\% slower than synthesis.

\noindent
\textbf{Cycles.} LightningSim produces cycle count results with 99.9\% accuracy on average, in comparison to the cycle counts produced by C/RTL cosimulation. LightningSim achieves 100\% accuracy for 26 out of 33 cases. In the remaining 7 cases, two have 1.9\% and 0.6\% error but both are only 1 cycle off; the other five are all within 0.2\% error.

\subsection{Parallelism with HLS Synthesis}

Fig.~\ref{fig:timelines} shows detailed timelines of the execution of LightningSim alongside HLS synthesis for a sample of benchmarks. LightningSim's efficiency advantage is clearest for large designs such as FlowGNN GIN, where each step of the synthesis process takes several minutes. Notably, in FlowGNN GIN, both trace generation and trace analysis are negligibly fast, while most of the time is spent idle waiting for HLS scheduling to finish.
For smaller designs, LightningSim also can often outperform synthesis, such as for Huffman encoding or vector addition with streams. 

\section{Conclusion}

In this work, we proposed LightningSim, a fast, accurate trace-based simulator for HLS. LightningSim maps the scheduling information produced during HLS synthesis onto the execution of HLS-generated LLVM bitcode. It does this in two stages, trace generation and trace analysis, which affords it unique flexibility enabling useful features, such as parallelizability with HLS synthesis and the ability to recalculate timings with different FIFO depths without re-simulating.

LightningSim leaves several directions for future work. Certain HLS features are not currently supported, for example, pipeline rewind and non-deterministic constructs. %
Additionally, a more accurate AXI model will enable LightningSim to achieve cycle-accurate simulation across more HLS designs.

LightningSim represents a step in the direction of agile hardware development, helping to bridge the Developer Experience~(DX) gap between software engineering and hardware design, and democratizing accelerator design, empowering more people to design specialized hardware confidently by enabling faster design iterations.

\section*{Acknowledgements}

This work is partially supported by the Center for Research into Novel Computing Hierarchies (CRNCH) at Georgia Tech and the 2022 Qualcomm Innovation Fellowship program.

\bibliographystyle{IEEEtranS}
\bibliography{refs.bib}

\end{document}